\documentclass[journal=jacsat,manuscript=article]{achemso}

\usepackage[version=3]{mhchem} 
\usepackage[T1]{fontenc}       
\usepackage{amsmath}
\usepackage{color}
\setcitestyle{numbers}%
\setcitestyle{square}
\author{Isha Malhotra\textsuperscript{1}}
\author{Sujin B. Babu\textsuperscript{1}}
\email{sujin@physics.iitd.ac.in}
\affiliation[Indian Institute of Technology]
{\textsuperscript{1}Department of Physics, Indian Institute of Technology, Hauz Khas, New  Delhi-110016, India}

\title[An \textsf{achemso} demo]
  {Aggregation kinetics of irreversible patches coupled with reversible isotropic interaction leading to chains, bundles and globules}

\begin{document}

\begin{abstract}
  In the present study we are performing simulation of simple model of two patch colloidal particles undergoing irreversible diffusion limited cluster aggregation using patchy Brownian cluster dynamics. In addition to the irreversible aggregation of patches, the spheres are coupled with isotropic reversible aggregation through the Kern-Frenkel potential. Due to the presence of anisotropic and isotropic potential we have also defined $3$ different kinds of clusters formed due to anisotropic potential and isotropic potential only as well as both the potentials together. We have investigated the effect of patch size on self-assembly under different solvent qualities for various volume fractions. We will show that at low volume fractions during aggregation process, we end up in a chain conformation for smaller patch size while in a globular conformation for bigger patch size. We also observed a chain to bundle transformation depending on the attractive interaction strength between the chains or in other words depending on the quality of the solvent. We will also show that bundling process is very similar to nucleation and growth phenomena observed in colloidal system with short range attraction. We have also studied the bond angle distribution for this system, where for small patches only $2$ angles are more probable indicating chain formation, while for bundling at very low volume fraction a tail is developed in the distribution. While for the case of higher patch angle this distribution is broad compared to the case of low patch angles showing we have a more globular conformation. We are also proposing a model for the formation of bundles which are similar to amyloid fibers using two patch colloidal particles.
\end{abstract}

\section{Introduction}\label{intro}

Biological particles self organize into highly monodisperse structures due to the presence of specific interaction sites on their surface \cite {whitesides2002self}; examples include virus, proteins etc \cite{bancroft1967study, salunke1989polymorphism, rombaut1990new, prevelige1993nucleation, mateu2013assembly}.  There are some specific kinds of proteins which aggregate together in ordered bundles leading to diseases such as cataract, Alzheimer's disease and Parkinson's disease \cite{chiti2006protein}. In fact it has been shown that almost all neurodegenerative diseases occur due to the abnormal protein aggregation. These kinds of proteins are generally termed as amyloid proteins \cite{woodard2014gel}. This type of bundling is also observed in semi-flexible polymers grafted on a surface \cite{benetatos2013bundling}. In this particular case an attraction between the polymer chains leads to the collapse of the homogeneous phase to a bundled state.   

To mimic some of these biological structures, colloidal particles which are asymmetric, patterned or patchy \cite{glotzer2007dimensions} have received considerable attention in recent years as their isotropic \cite{dorsaz2011phase, valadez2012phase, abramo2012effective} counterpart was not able to explain some of the experimental observations like bundling. Particles which are anisotropic in shape or interactions are synthesized as they have very promising applications in electronics\cite{gratzel2001photoelectrochemical} , drug delivery \cite{langer2004designing, champion2007making}, in fabricating photonic and plasmonic materials \cite{liddell2003stereological, zhang2005self, halverson2013dna, coluzza2013design, liu2016diamond} . Several patchy particles have been developed where colloidal particles undergo DNA-mediated interactions and are known as DNA-functionalised colloids \cite{wang2012colloids, feng2013dna, di2013developments} . The patchy models have been already used to study the equilibrium properties of polymerization  by Wertheim theory \cite{wertheim1987thermodynamic} where they developed a thermodynamic perturbation theory for patchy particles having two patch sites. Sciortino et al. \cite{sciortino2007self} did an extensive study of the two patch site model and have shown that simulation matches with the predictions of Wertheim theory. In their study they have looked at the equilibrium structures formed when only one bond per patch was allowed. Structures such as tubes, lamellae are predicted in experimental and computer simulations by varying size and shape of the patches \cite{chen2011supracolloidal, munao2013cluster, preisler2013phase, vissers2014cooperative, preisler2014equilibrium} . Although extensive work has been carried out for the case of patchy particles and inverse patchy colloidal particles \cite{bianchi2006phase, foffi2007possibility, bianchi2008theoretical, bianchi2011patchy, noya2014phase, kalyuzhnyi2015theoretical, ferrari2015inverse, noya2015phase, wolters2015self, hatch2015computational, ferrari2017spontaneous} , very few work has been carried out for the case where in addition to patchy interaction an isotropic potential is also considered along with it\cite{dudowicz2003lattice, rah2006lattice, liu2007vapor,dudowicz2009exactly, li2009simple,audus2016coupling}. These studies were confined to finding the equilibrium properties like liquid-liquid coexistence curve or to study the competition between phase separation and polymerization. Liu et al. \cite{ liu2007vapor} have observed that as the patch number decreases the critical point of the liquid-liquid binodal curve shifts towards the lower volume fraction and width of the binodal curve increases. Using these results they were able to explain the phase separation of proteins like lysozyme and $\alpha-$crystalline molecule. Dudowicz and coworkers \cite{dudowicz2003lattice, rah2006lattice, dudowicz2009exactly} studied the competition between phase separation and polymerization using lattice-based linear polymerization models. Audus et al. \cite{audus2016coupling} recently studied the structural property of the equilibrium structure in a similar system, where they showed that the fractal dimension of this system turns out to be $2$. It has already been shown that the reversible system is close to reaction limited aggregation model in the case of only isotropic potential and a fractal dimension of $2$ \cite{babu2006flocculation}. M. Kastelic et al. \cite{kastelic2015protein} also studied a similar system where they estimated the number of patches and interaction strength of the patch using experimental data to reproduce the liquid-liquid coexistence curve, but in their study isotropic potential was not considered. In the present work, we have modeled a unique system where the bonds between the patches are irreversible while the bonds formed via isotropic potential are reversible. When two patches come in close contact they form a bond or in other words it follows diffusion limited cluster aggregation model \cite{babu2006flocculation}, which has a fractal dimension of $1.8$. As  Prabhu et al. \cite{prabhu2014brownian} have shown for monovalent patches, particles organize themselves into chains(fibers) and these chains form an arrested structure which consists of strands. Formation of fibers and bundles has also been observed by Huisman et al. \cite{huisman2008phase}  where they used an anisotropic Lennard-Jones potential to model two-patch particles. In this system above a particular temperature these particles form bundles, and they have shown that this transition is very similar to the sublimation transition for polymers. Also a transition from small clusters to long straight rigid tubes has been observed by Vissers et al. \cite{vissers2014cooperative} in case of one-patch colloid of 30$\%$ surface coverage below a specific temperature which is density dependent. 

In the present work we are using the simulation technique called the Brownian cluster dynamics(BCD). This method was developed for studying the structure, kinetics and diffusion of aggregating system of particles interacting via isotropic potential \cite{babu2008influence, babu2009crystallization, shireen2017lattice}. It has already been shown that BCD agrees with other Brownian dynamics simulations \cite{rottereau2005influence} and predicted phase diagrams are similar to Monte Carlo simulations \cite{babu2008influence, babu2006phase, babu2009crystallization} . The advantage over the other methods is that BCD can handle very large number of particles upto $10^6$ particles. Acutha et al. \cite{prabhu2014brownian} modified this method to accommodate asymmetric potential with isotropic potential on the particles. They have shown that when a single polymer chain was simulated using this technique they were able to get the correct static and dynamic properties ignoring hydrodynamic interactions. They went further and modeled the step growth polymerization where one patch could form only one irreversible bond under different solvent conditions where they observed that the kinetically driven system formed an arrested structure. In the present work we do not have any constraint over the number of bonds a patch can form, which leads to quite fascinating structures.

The paper is arranged in the following ways. In model and simulation techniques we briefly describe our simulation technique and also explain how we are changing the quality of the solvent. Then we present our results about the change in the structure of the system as we vary the patch size. For very small patches we form linear chains  while on increasing the patch size we end up in a more globular structure, also the effect of changing the volume fraction of the system will be discussed. We will also discuss about the possible link between protein and patchy particle aggregation in the context of bundling, which is believed to be the origin of many neurodegenerative diseases, followed by conclusion in last section.

\section{MODEL AND SIMULATION TECHNIQUES}\label{modeltechniques}
A model potential was developed by Kern and Frenkel \cite{kern2003fluid} to simulate colloidal systems with anisotropic interactions. This model consists of hard spheres of diameter $\sigma$ which is kept as unity with two patches, where the orientation of the patches is specified by a unit vector $\hat{\bf v}_{i}$,  which lies at the center of patch.  The patch can be viewed as intersection of sphere with a cone of solid angle $2 \omega$ having vertex at the center of the sphere. We simulate hard spheres complemented with patch vectors $\bf {v}_{i}$ interacting through a square well patchy potential ($U_{iso} (r_{i,j})$) followed by Kern Frenkel potential ($U_{patchy} ({\textbf{\^r}}_{i,j},{\textbf{\^v}}_{i},{\textbf{\^v}}_{j})$). Here ${\hat{\bf r}}_{i,j} = {\bf r}_{i,j}/r_{i,j}$, where ${\bf {r_{i,j}}}$ is the vector connecting the center of mass of spheres $i$ and $j$,  ${\bf{\hat{v}}}_{i}={\bf v}_{i}/{v_{i}}$,  ${\bf{\hat{v}}}_{j} =  {\bf v}_{j}/{v_{j}}$ are unit patch vectors of spheres $i$ and $j$. The potential is given by
 
\begin{equation}
 U_{tot} (\mathbf{r}_{i,j},  {\bf v}_{i} , {\bf v}_{j})  = 
 \begin{cases}
   \infty\hspace{2.2cm} r_{i,j}\leq \sigma 
   \\
   U_{iso} + U_{patchy}\quad \sigma < r_{i,j}\leq\sigma(1+\epsilon)
\\
0\hspace{2.3cm}  r_{i,j}>\sigma(1+\epsilon)
\end{cases}
\label{e.1}
 \end{equation}
 \begin{equation}
 U_{iso} (r_{i,j}) = -u_0 
\label{e.2}
\end {equation}
where $\epsilon$ is the interaction width which we have kept $\epsilon=0.1$ and $u_0$ is the depth of the square well and $\sigma$ is the diameter which is kept as unity in the present study. $U_{patchy}$ depends on the orientation of the two particles and is defined as:
\begin{equation}
U_{patchy} ({\textbf{\^r}}_{i,j},{\textbf{\^v}}_{i},{\textbf{\^v}}_{j}) = 
\begin{cases}
-u_1\: \text{if}\: {\textbf{\^r}}_{i,j}.{\textbf{\^v}}_{i}>\text{cos}\ \omega\: \text{and}\:{\textbf{\^r}}_{j,i}.{\textbf{\^v}}_{j}>\text{cos}\ \omega \\
0\quad \text{else}
\end{cases}
\label{e.3}
\end{equation}
In the present study each patch vector is associated with two oppositely located patches and $\omega$ is a tunable parameter where we can change the percentage of the patchy surface coverage on the sphere. $\omega = 90^{\circ}$ corresponds to the irreversible aggregation of pure isotropic square-well potential and $\omega = 0^{\circ}$ corresponds to the hard sphere. 

We start our simulation with an ensemble of $N_{tot}$ randomly distributed spheres (each associated with a randomly oriented patch vector) of diameter $\sigma = 1$ in a box of length $L$, where volume fraction  $\phi=\pi/6 N_{tot}/L^3$. We have used a cubic box of fixed length $L = 50$ with periodic boundary conditions in the present study. The simulation procedure involves $2$ steps one is the cluster construction step and the other one is the movement step. In the cluster construction step, when $2$ monomers or spheres are within the interaction range then a bond is formed with a probability $\alpha_0$, while if a bond already exists then it is broken with a probability $\beta_0$, so that $P=\frac{\alpha_0}{\alpha_0+\beta_0}$. Then $P=1-\exp(u_0/k_BT)$ is the relation connecting temperature $T$ with the probability $P$ as already shown by Prabhu et al. \cite{prabhu2014brownian} . The collection of all bonded spheres together is defined as a cluster. For the square well system we have defined the reduced second virial coefficient \textit{$B_2=B_{rep} - B_{att}$}, where $B_{rep}=4$, which is the repulsive part coming from hard core repulsive interaction between the spheres. $B_{att}$ is the attractive part of the second virial coefficient due to the square well potential and is given by$B_{att} = 4.\ [ exp( - u_{0}/ k_B . T)]\ .\ [ (1+\epsilon)^3 - 1]$\cite{kihara1953virial}, where $T$ is the temperature and $k_B$ is the Boltzmann factor, which is kept as unity in the present work. When two patches overlap the probability to form a bond is $\alpha_p=1$ and the probability to break a bond is kept as $\beta_p=0$, thus forming irreversible bonds. In the present study we have allowed multiple bonds per patch which is quite different from the previous studies where they have employed an artificial constraint of single bond per patch. 
In the movement step we randomly select $2N_{tot}$ times a sphere then we either rotate it whereby we rotate the patch vector randomly with an angular displacement of $s_R$ with respect to the patch vector or we translate the sphere with a small step size $s_T$ in a random direction. Thus on average every particle undergoes both rotational and translational diffusion independently and in an uncorrelated manner  as already demonstrated by Prabhu et al.\cite{prabhu2014brownian} . If the rotation or the translation step leads to the breaking of bond or overlap with the other spheres that movement is rejected in our simulations. The step sizes have to be very small otherwise it will lead to a nonphysical movement due to rejection of the movement steps. It has already been shown for the parameters chosen in the present study, in order to obtain the correct diffusional behavior, the translation step size $s_T=0.013$ and the rotational step size $s_R=0.018$  are the best choices of parameters\cite{prabhu2014brownian}. After a cluster construction and movement step is over the simulation time $t_{sim}$ is incremented. The relation between physical time $t$ and $t_{sim}$ comes from the fact that for a free particle undergoing Brownian motion, mean squared displacement is given by $<R^2>=6D_1^T t$, where $D_1^T =1/6$ is the translational diffusion coefficient of a single sphere. For a particle undergoing random walk $<R^2>=t_{sim} s_T^2$ where $s_T$ is a constant step size and $t_{sim}$ is the number of simulation steps taken.  The time taken for a sphere to travel its own diameter $\sigma$ is given by $t_0= <\sigma^2>$ therfore the reduced time is given by $t/t_0=t_{sim} s^2_T$.

In the present work we have used three different $B_{att}$ values:
\begin{itemize}
  \item $B_{att} = 0$ : In this case system is in good solvent condition, there is no reversible isotropic interaction and the particles aggregate only due to  irreversible bonding between the patches.
  
  \item $B_{att} = 4$ : As $B_{att}$ increases, quality of solvent deteriorates or in other words reversible isotropic aggregation also contributes towards the structure. At this value of $B_{att}$ the hard core repulsion is balanced by the attractive part of the potential and $B_{2}=0$. This condition corresponds to Boyle temperature of the fluid.
   
\item $B_{att}=12$ : For pure isotropic fluid, this $B_{att}$ corresponds to the value where we observe liquid-crystal binodal \cite{babu2009crystallization} or can be considered as bad solvent condition.
\end{itemize}

\begin{figure}
\includegraphics[scale=0.5]{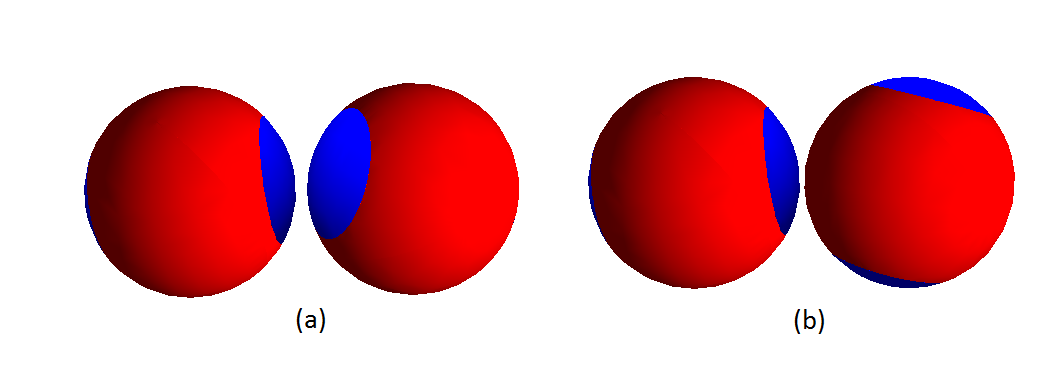}
  \caption{The red surface of sphere is hard sphere which undergoes only isotropic interaction and blue surface are the patches which undergoes anisotropic interaction (a) Monomers forming irreversible bonds through patches ($P$ type clusters). (b) Monomers interacting through reversible isotropic interaction ($NPI$ type clusters).}
  \label{fig1}
\end{figure}

 We wil be working with the above three \textit{$B_{att}$} conditions, mainly for $2$ different  volume fractions $\phi=0.02$ and $\phi=0.2$ and  for two different $\omega$ values $22.5^{\circ}$ and $45^{\circ}$.  As a result of irreversible patchy and reversible isotropic interaction we are able to define $3$ different types of clusters (Fig.  \ref{fig1}), 
\begin{itemize}
\item We define spheres which interact only through the patches, or spheres which are irreversibly connected to its neighbors as $P$ cluster as shown in figure \ref{fig1}(a).
\item The spheres which only interact through the isotropic potential forms a different type of cluster which we call as $NPI$ cluster see figure \ref{fig1}(b).
\item The clusters  formed as a result of both patchy as well as isotropic interaction are called as $PI$ cluster.
\end{itemize} 
In the present work we run our simulation till the system forms a a percolating cluster in $PI$ construction i.e. cluster that extends from one end of the box to the opposite end.

\section{Results and discussion}\label{result}
\begin{figure}
\includegraphics[scale=0.5]{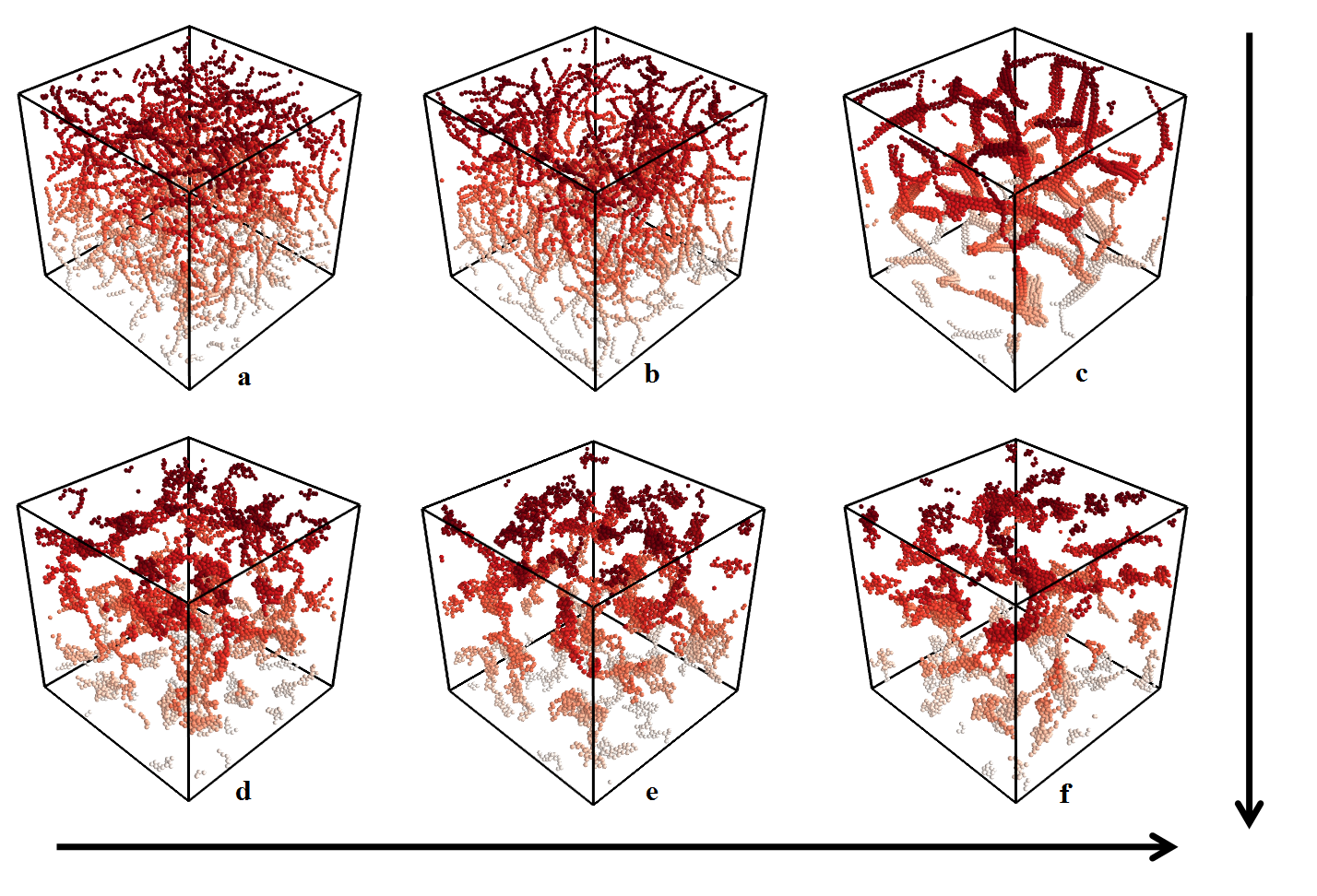}
  \caption{Snapshots of the system at same physical time $t/t_0=1800$. From left to right, $B_{att}$ is increasing from $B_{att}$=0, 4, 12 and the top picture is at $\omega = 22.5^{\circ}$ and bottom is at $\omega = 45^{\circ}$. }
  \label{structures}
\end{figure}
In the present work we have used two different patch angles $\omega = 22.5^{\circ}$ and $\omega = 45^{\circ}$ to study the aggregation of patchy particles mainly at two different volume fractions $\phi= 0.02$ and $\phi=0.2$. The effect of the solvent condition was taken into account by changing the $B_{att}$ as mentioned in previous section. In  Fig. \ref{structures} we are showing the structures obtained from the simulations at $B_{att}=0$, $B_{att}=4$ and $B_{att}=12$ at two different $\omega$ values $22.5^{\circ}$ and $45^{\circ}$ for $\phi=0.02$ after the same time $t/t_0=1800$, where the kinetics of the system no longer evolves. In fig \ref{structures}(a) $B_{att}=0$ and $\omega=22.5^{\circ}$, we observe the formation of chains. As the patch angle is very small  it is not possible to form more than one bond per patch due to hard core repulsion between the spheres, reversible isotropic potential also does not play any part in the aggregation process as $B_{att}=0$, so only chain formation is possible. In fig \ref{structures}(b) we show the same system for $B_{att}=4$, a visual inspection itself reveals that the length of the chains is larger compared to $B_{att}=0$. This is due to the presence of isotropic attractive potential between the spheres, they stay in each others range for a longer period of time compared to $B_{att}=0$. As the particles diffuse within the bonds as well as the spheres also undergo rotational diffusion they are more likely to form irreversible bond through patches. For the case of $B_{att}=12$ we observe that the reversible part of the potential plays a major role which results in the transformation from chains to bundles as can be observed in fig \ref{structures}(c). In fig \ref{structures}(d) we have shown the structure for $B_{att}=0$ for $\omega=45^{\circ}$ where we observe the presence of dense clusters, as spheres are able to form multiple bonds per patch. For the case of $B_{att}=4$, the number of bonds per particle will increase as the reversible aggregation will also contribute towards the structure formation and we observe denser clusters. In the case of $B_{att}=12$ the clusters should have been more denser, which we do not observe in the fig \ref{structures}(f) as the dominant contribution for aggregation comes from the irreversible part of the potential. Irreversible aggregation always leads to branching and forms fractal type of aggregate.
\begin{figure}
\includegraphics[height=9cm,width=9cm]{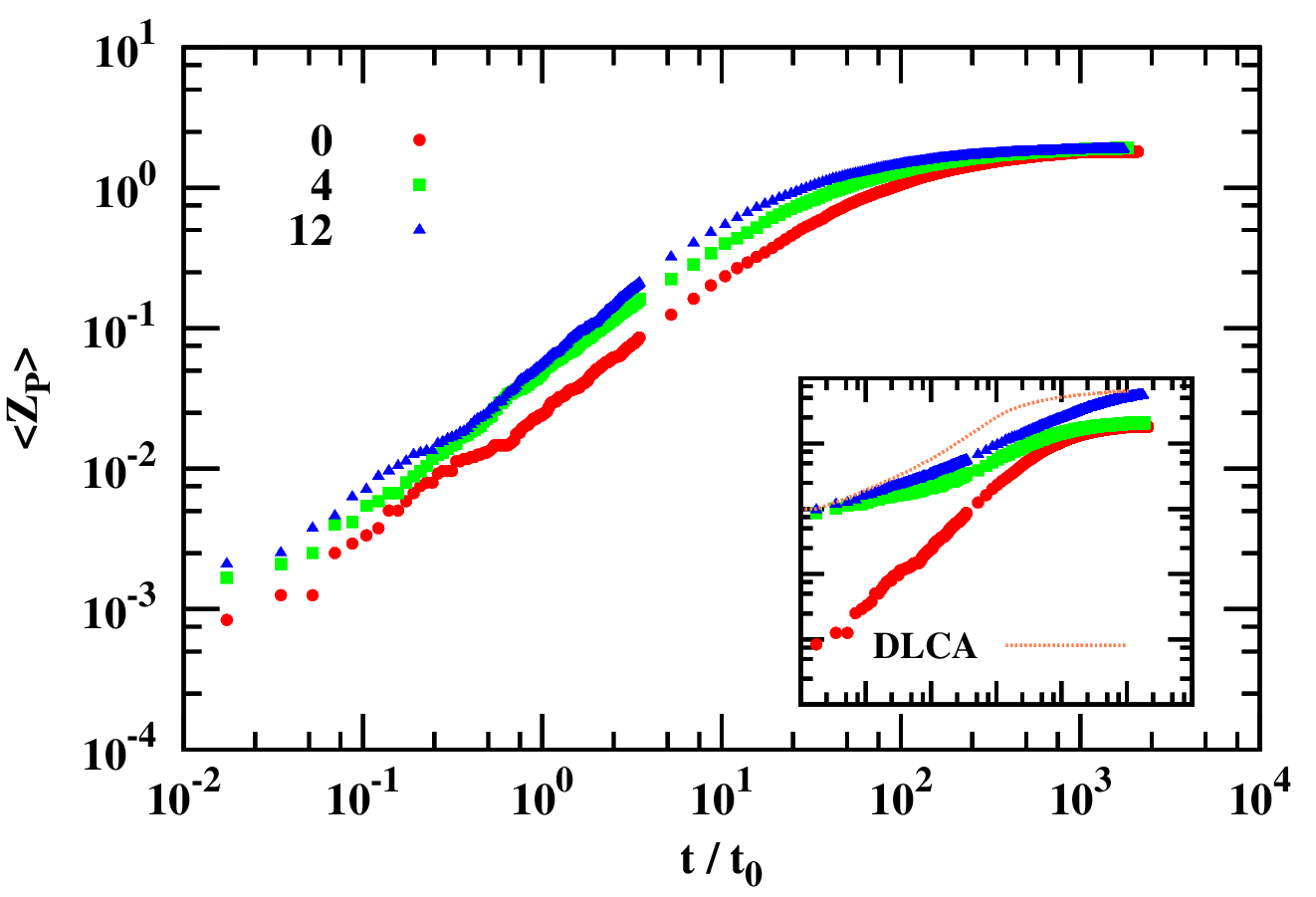}
  \caption{The average bonded neighbors for $P$ type cluster $<Z_{P}>$  is plotted with respect to reduced time at different $B_{att}$ values as indicated in the figure for $\phi=0.02$ at $\omega=22.5^{\circ}$. The inset shows average bonded neighbors for $PI$ type cluster$<Z_{PI}>$ where the solid line is for irreversible DLCA system for the case of pure isotropic square well potential.}
  \label{fig2}
\end{figure}
\subsection{Kinetics of aggregation}\label{kinetics}
In order to understand the kinetics of aggregation  we have followed the average number of bonded neighbors $Z$ as a function of time for three different cluster construction type as explained in previous section. In Fig. \ref{fig2} we have plotted the average number of bonded neighbor for the $P$ construction (clusters formed only due to the patches) at $\phi=0.02$ for different $B_{att}$ values. The aggregation process starts from a randomly distributed spheres system and thus for all the three $B_{att}$ we start at the same value of $Z_p$ as the bonds are irreversible for the patches. If the distance between the center of mass of the sphere is within a distance of $1+\epsilon$ and also the patches are aligned irreversible bonds are formed between the patches, while if patches are not aligned but are within the interaction range a reversible bond is formed. The average life time of such a reversible bond is given by $1/\beta$, so as $B_{att}$ increases the life time of the bond increases which means the sphere will be in each other's interaction range for a longer period of time. The spheres stay in each other's interaction range for a longer time and thus the patch vectors will align themselves thus forming an irreversible bond. As a result in the $P$ construction the kinetics of aggregation becomes faster as we increase the attractive strength. We observe that for $B_{att}=0$, the number of bonded neighbors for $P$ construction stagnates at a value $\sim 2$, which means for $\omega=22.5^{\circ}$ on average only $2$ bonds per particle are formed or in other words a patch is able to form only one bond.  In the inset  of Fig. \ref{fig2}  we have plotted the average number of neighbors which are connected to a sphere through both irreversible as well as reversible part of the potential ($PI$ construction). For the case of $B_{att}=0$ and $B_{att}=4$ the average number of neighbors reaches a steady state value close to $2$ signifying there is no densification in these systems. As the $Z_{PI} \sim 2$ also signifies that  for the $PI$ construction as well, we have on average $2$ bonds per particle and our system will have only chain like configuration for smaller values of $B_{att}$. For the case of $B_{att}=12$ we observe that the average number of neighbors increases to a value more than $2$, indicating that we have a locally more dense system. This is expected because pure isotropic square well (system where we do not have anisotropic patchy interaction) counterpart undergoes phase separation through nucleation and growth phenomena at $B_{att}=12$ \cite{babu2009crystallization}. For comparison we have also plotted the case when we have pure square well potential undergoing irreversible DLCA \cite{babu2008diffusion} shown as dotted lines in the inset of the Fig. \ref{fig2}. Here we observe that the kinetics of aggregation for the case of irreversible pure isotropic square well and patchy particles are very different from each other indicating that isotropic reversible part of the potential is playing a major role in the aggregation phenomena as well as the structure. 

\begin{figure}
\includegraphics[height=9cm,width=9cm]{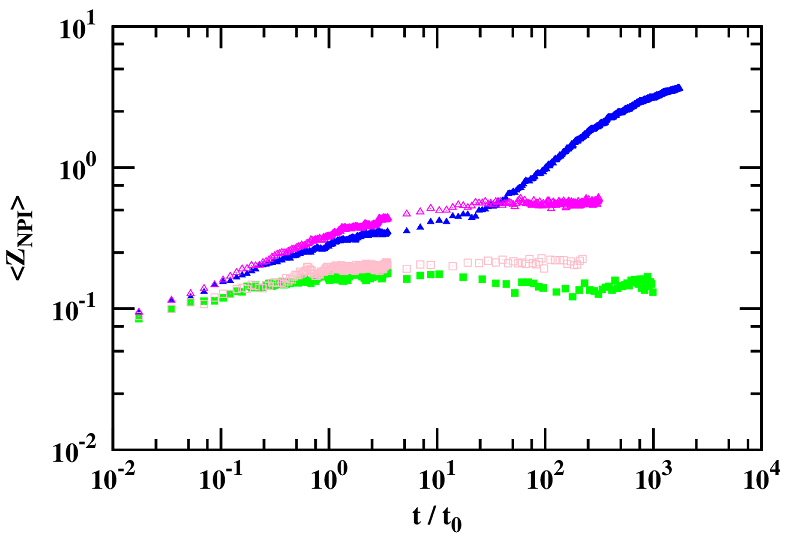}
  \caption{The evolution of $<Z_{NPI}>$ is plotted with respect to reduced time  at $\omega=22.5^{\circ}$ for $\phi=0.02$ for $B_{att}=4$ (square) and $B_{att}=12$ (triangles). The filled symbols indicate irreversible patchy particles with reversible isotropic interaction and open symbols indicate reversible aggregation of purely isotropic square well potential.}
  \label{fig3}
\end{figure}

To understand the role played by the reversible part of the potential we have plotted in Fig. \ref{fig3} the evolution of the average number of neighbors ($Z_{NPI}$) formed due to only the reversible part of the potential as a function of time for $\phi=0.02$ and $\omega=22.5^{\circ}$. For $B_{att}=0$, we have $\alpha_0=0$, thereby no reversible bonds are formed, while for $B_{att}=4$ and $12$ we have both $\alpha_0>0$ and $\beta_0>0$. For the case of $B_{att}=4$, the contribution of $Z_{NPI}$ towards $Z_{PI}$ is more compared to $Z_P$ in initial time of the aggregation process $t/t_0<0.2$, which means the initial aggregation process is dominated by the reversible part of the potential and at later times attains a constant value {\color{red}0.14}. In Fig. \ref{fig3} we also observe that the kinetics of aggregation of the reversible part of the patchy particle and pure isotropic square well potential are very similar, although for the patchy case the curve is always below the pure square well case as in this calculation we do not consider the particles that are bonded by irreversible bonds. After a time $t/t_0>10$, $Z_{NPI}$ starts to decrease slightly as more and more particles are forming part of the $P$ cluster as evident from Fig. \ref{fig2}, where $Z_P$ keeps on increasing which means aggregation process is dominated by the irreversible aggregation of the patchy sites. For the case of $B_{att}=12$ we observe that $Z_{NPI}$ follows similar trend as that of $B_{att}=4$ although the average number of neighbors are higher than $B_{att}=4$ as the attraction strength in the case of former is higher than latter. We also observe that after a time of $t/t_0>30$, the kinetics of aggregation is accelerated for the case of patchy particle at $B_{att}=12$, indicating a chain to bundle transition. For the case of pure isotropic potential this upturn indicates the gas-crystal phase separation which we do not observe in the case of $Z_{NPI}$ as the system is stuck in the meta stable state as shown by Babu et al. \cite{babu2009crystallization}. For the case of patchy particle we do not observe any kind of meta stable state for chain to bundle transition.
\begin{figure}
\includegraphics[height=9cm,width=18cm]{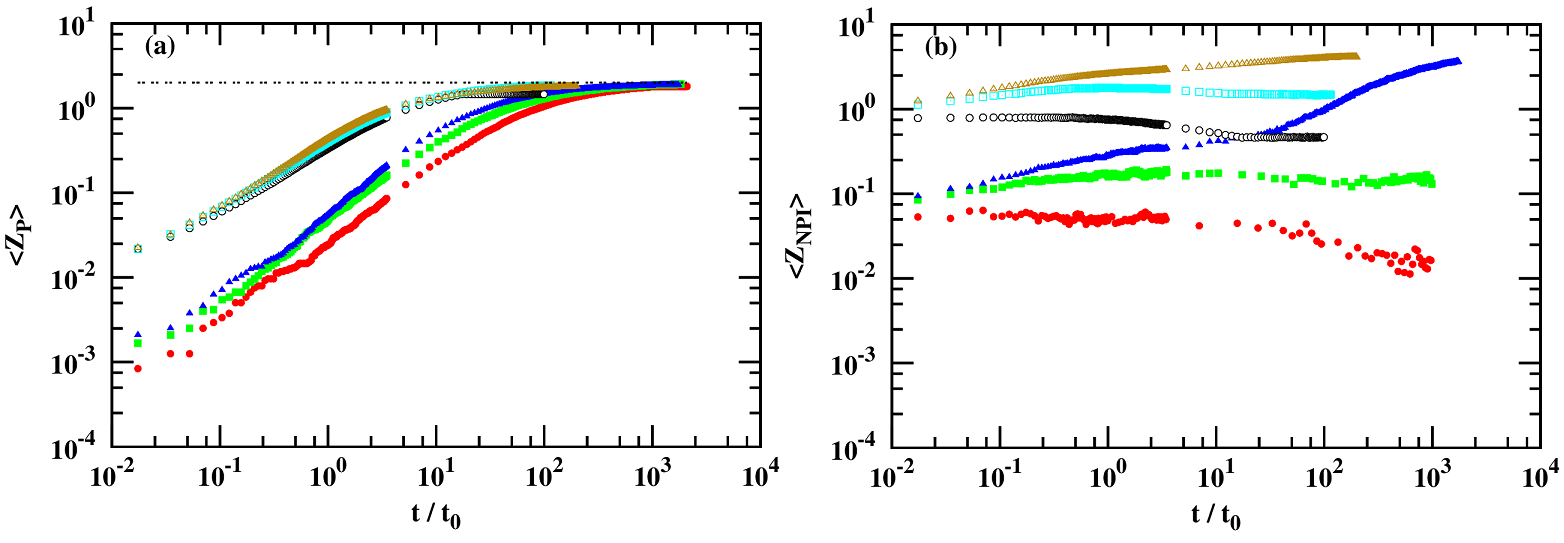}
  \caption{Comparison between the average bonded neighbors for $\phi=0.02$ (closed symbol) and $\phi=0.2$ (open symbol) at $B_{att}=0$ (circle), $B_{att}=4$ (square) and $B_{att}=12$ (triangle) for $\omega=22.5^{\circ}$. (a). The average number of bonded neighbor for $P$ type cluster $<Z_{P}>$  is plotted as function of reduced time $t/t_0$. (b). The average number of bonded neighbor for $NPI$ type cluster$<Z_{NPI}>$  is plotted as function of reduced time $t/t_0$.The double dotted line indicates the value when the average number of bonded neighbor is $2$. }
  \label{fig4}
\end{figure}

As observed in the irreversible aggregation of pure square well fluids, the volume fraction of the system also plays a major role in deciding the kinetics and structure of the patchy particle aggregate \cite{babu2008diffusion}. In figure \ref{fig4}(a) we have plotted the evolution of the average number of neighbors due to the patchy part of the potential $Z_P$ as a function of time for $\phi=0.02$ and $\phi=0.2$ at $\omega=22.5^{\circ}$. As we increase the volume fraction of the system the number of particles with their bond vector oriented against each other as well as the number of particles within each other's range increases. This is the reason why $Z_P$ starts at a higher value for $\phi=0.2$ compared to $\phi=0.02$ where we have started both the aggregation process from a random system. For the case of $P$ construction we observe that $Z_P \sim 2$ indicating the fact that for higher $\phi$ also on average the patches are able to form only one bond per patch when $\omega=22.5^{\circ}$. The value of $Z_P$ also reaches a steady state value of $ \sim 2$ for $B_{att}=4$ and $B_{att}=12$ but is slightly above for $B_{att}=0$ indicating the fact that more number of patches have formed bonds thereby the clusters have increased in size. For the case of $NPI$ construction see figure \ref{fig4}(b) the number of neighbors $Z_{NPI}$ for $B_{att}=4$ rises initially and then goes down before attaining a steady state value quite similar to the case of $\phi=0.02$. When the spheres are in each other's interaction range, they may form a reversible bond and then they diffuse within the bonds such that their patches get aligned whereby they will form irreversible patchy bond, thus the number of reversible bonds reduces. In the case of higher volume fraction and weaker reversible attraction  $B_{att}=4$, we form chains and the chain length is higher as compared to the case of $B_{att}=12$ which forms bundles. For the case of $B_{att}=12$ even though the upturn in $Z_{NPI}$ is not as evident as in the case of $\phi=0.02$, still this system tries to densify but it is hindered as the movement of the spheres are restricted due to the increased number of particles in the system.
  \begin{figure}
\includegraphics[height=9cm,width=9cm]{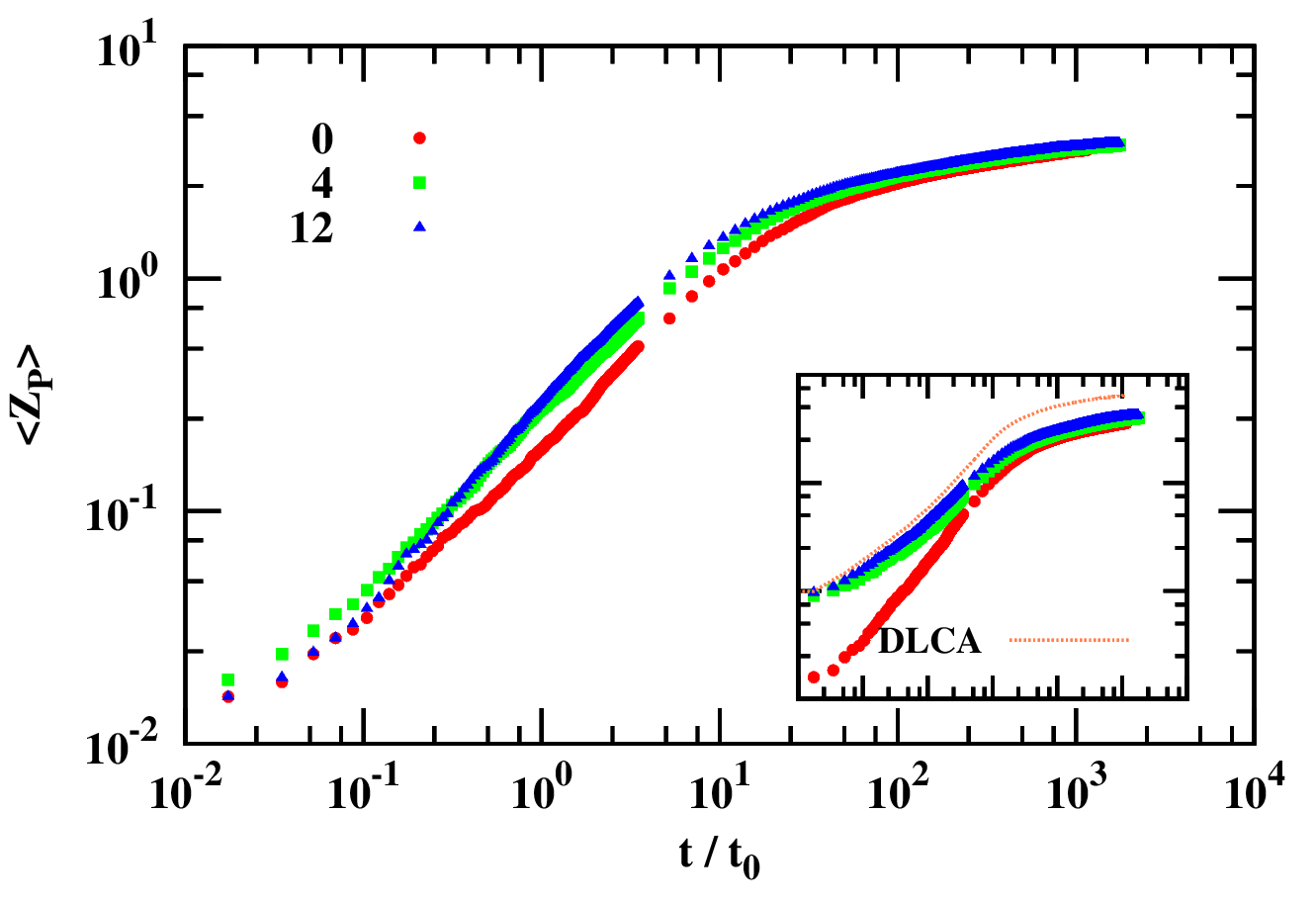}
  \caption{$<Z_{P}>$ is plotted as a function of reduced time for $\phi=0.02$ at  $\omega=45^{\circ}$ when $B_{att}=0$ (circle), $B_{att}=4$ (square) and $B_{att}=12$ (triangle). The inset shows the average number of neighbors for $PI$ type cluster and the dotted line indicates the irreversible DLCA aggregation when only isotropic interaction is present.}
  \label{fig5}
\end{figure}

As explained in previous section in our simulation we are allowing multiple bonds per patch, which means the patch angle will also play a significant role in the kinetics as well as the structure that is formed during the aggregation process. In Fig. \ref{fig5} we have plotted the $Z_P$ as a function of the reduced time for the case $\omega=45^{\circ}$ for $\phi=0.02$. $B_{att}=0$ has the slower kinetics and attains a steady state value for $t/t_0>50$. As expected $B_{att}=12$ has higher number of patchy neighbors compared to the other $B_{att}$ values, as the spheres are connected through the reversible part of the potential and hence they are more likely to diffuse within bonds to form irreversible patchy bonds. In all the three cases $Z_P$ attains a steady state value greater than $2$ indicating multiple bonds are formed per patch. As $Z_P>2$ we are not observing any chain formation for higher $\omega$ value but more globular type structure. In the inset we have shown the evolution of $Z_{PI}$ as a function of time, in all the cases it goes to the same steady state value which means that for higher patch angles the irreversible part of the potential dominates compared to $\omega=22.5^{\circ}$. For earlier time $Z_{PI}$ for $B_{att}=4$ and $B_{att}=12$ starts from a higher value compared to $B_{att}=0$ as the number of bonded neighbors increases due to the reversible interaction of the potential. For the case of $B_{att}>0$ the value of $Z_{PI}$ will always be smaller than for the case of pure isotropic irreversible  DLCA. Although the evolution follows very closely the pure isotropic DLCA, indicating that system wants to evolve the same way, but is restricted due to the finite patch size. On increasing $\omega$ further we will converge towards the  irreversible pure isotropic DLCA system \cite{babu2008diffusion}. 

\begin{figure}
\includegraphics[height=9cm,width=9cm]{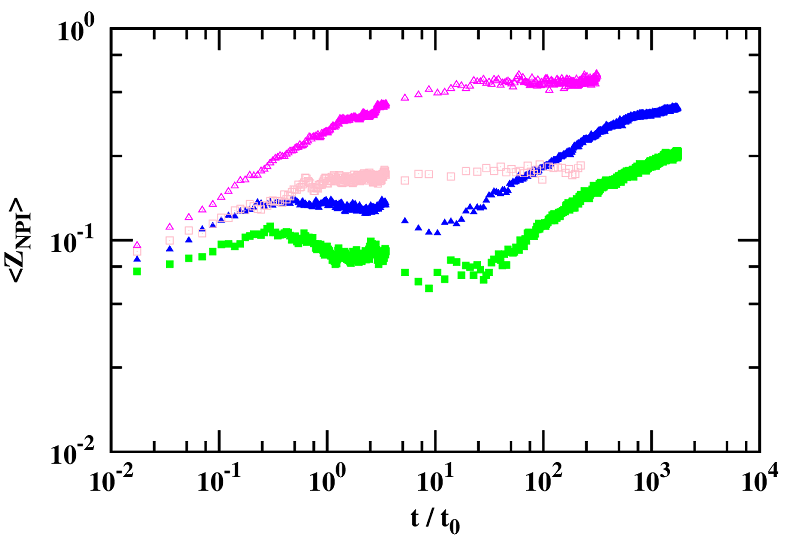}
  \caption{$<Z_{NPI}>$ is plotted as a function of time $t/t_0$ for $\phi=0.02$ when the value of  $\omega=45^{\circ}$ at $B_{att}=4$ (square) and $B_{att}=12$ (triangle). The filled symbols indicate the system with both irreversible patchy and reversible isotropic interaction and open symbols indicate reversible purely isotropic square well potential. }
  \label{fig6}
\end{figure}

In Fig. \ref{fig6} we have plotted $Z_{NPI}$ as a function of reduced time for $\phi=0.02$ for the case of  a pure square well reversibly aggregating system and only the reversible part of the patchy particle with $\omega=45^{\circ}$. We observe that in the initial times the contribution of the isotropic part of the potential is not significant compared to the irreversible aggregation of the patchy particles as the maximum steady state value is less than $0.9$, which means on average  less than one particle is reversibly connected to each particle. The particles which were connected through the reversible part of the potential later diffuses and forms irreversible patchy bonds, which is the reason for the dip we are observing in $Z_{NPI}$. These small clusters formed will diffuse and further densify into a globular form thereby $Z_{NPI}$ starts to increase even though the  rate of increase is very slow. The increase in $Z_{NPI}$ for $\omega=45^{\circ}$ is characterized by the increase in the size of the clusters as observed in figure \ref{fig2}(e), whereas for $\omega=22.5^{\circ}$ the sudden increase in $Z_{NPI}$ leads to chain to bundle transition.
\begin{figure}
\includegraphics[height=9cm,width=9cm]{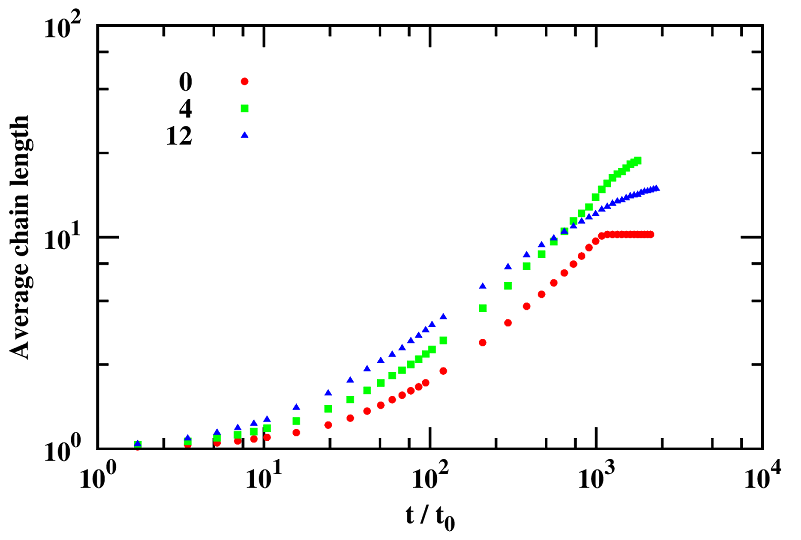}
  \caption{The growth of average chain length as a function of time at $\omega=22.5$ and $\phi=0.02$ for different $B_{att}$ as indicated in the figure.}
  \label{fig7}
\end{figure}

We have already shown that for $\omega=22.5^{\circ}$ chains are formed irrespective of the $B_{att}$ used in the system. In Fig. \ref{fig7} we have plotted the chain length as a function of time for a volume fraction $\phi=0.02$. For calculating the chain length we identify the sphere with only one bonded neighbor through the patches, the neighboring sphere will have $2$ neighbors through the patches if it is connected to a chain. We keep on counting the connected neighbor through the patches till we reach a monomer with only one bond, which will be the opposite end of the chain. The average chain length is defined as $<l>=\sum m_l N(m_l)/\sum N(m_l)$, where $m_l$ is the mass of the chain and $N(m_l)$ is the size distribution of chain lengths. For the case of $B_{att}=0$ the average chain length reaches a constant value $9.7$, which is consistent with the predictions of Sciortino et al. \cite{sciortino2007self} . While for the case of $B_{att}>0$ studied in the system the chain length goes on increasing, indicating that as the reversible interaction of the particles increases the length of the chain is larger at the initial times, as can be observed when we compare the Fig. \ref{fig2} and Fig. \ref{fig3}. At longer times we observe that the chain length for $B_{att}=4$ crosses over $B_{att}=12$, i.e on average we have longer chains for the case of smaller attraction. This may seem contour intuitive, but what we observe for $B_{att}=12$, there is a chain to bundle transition, while for $B_{att}=4$ the system remains as chains. For $B_{att}=4$ the particles or chains get attached to the other chains and then they  have enough time either to break away or diffuse to form irreversible patchy bonds thereby increasing the chain length. For $B_{att}=12$ we observe similar kinetics but as the attraction is higher between the particles, once they form part of the chain they start to aggregate as bundles.     

\subsection{Structural Analysis}\label{structure}
\graphicspath { {/} }
\begin{figure}[h!]
\centering
\includegraphics[height=9cm,width=9cm]{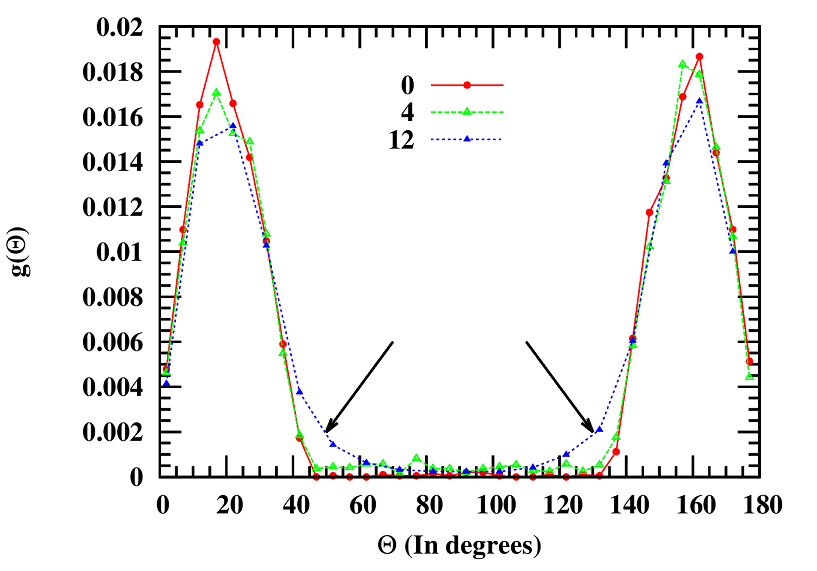}
\caption{$g(\theta)$  the probability distribution of the angle for a single sphere due to the bonded particle is plotted at different $B_{att}$ as indicated at $\omega=22.5^{\circ}$. The arrows indicate the extended tail of the distribution indicating a chain to bundle transition.}
\label{fig10}
\end{figure}
\begin{figure}[h!]
\centering
\includegraphics[height=9cm,width=9cm]{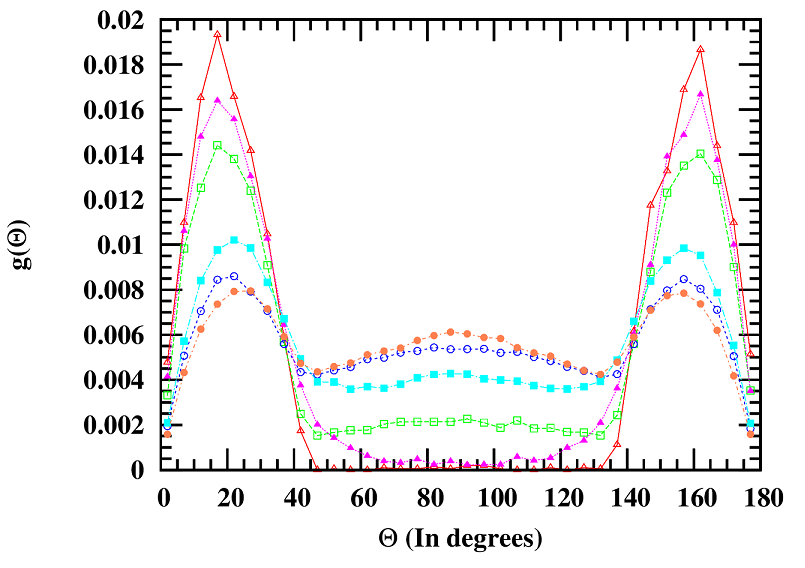}
\caption{The probability distribution $g(\theta)$ is plotted for $B_{att}=0$ (open symbols)  and $B_{att}=12$ (closed symbols) at $\omega=22.5^{\circ}$ for $\phi=0.02$ (triangle), $\phi=0.2$ (square) and $\phi=0.4$ (circle).}
\label{fig11}
\end{figure}
\begin{figure}[h!]
\centering
\includegraphics[height=9cm,width=9cm]{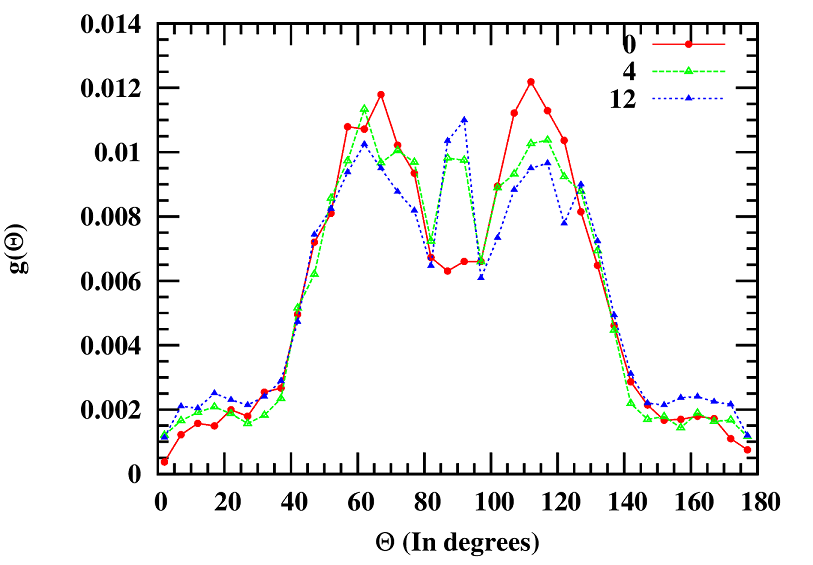}
\caption{The probability distribution$g(\theta)$ is plotted for $\phi=0.02$  at $\omega=45^{\circ}$ for a range of $B_{att}$ as indicated.}
\label{fig12}
\end{figure}

To identify where the spheres are distributed in the patches we have calculated the probability of occurrence of an angle for a single particle $g(\theta)$ between the patch vectors of two particles (the area under the curve has been normalized to one). If both the patches face in the same direction then the angle between the patch vectors is zero while if they face each other the angle between them is defined as $180^{\circ}$. The distribution of the angle defined between all the patch vectors in the system $g(\theta)$ is quite similar to the  pair correlation function where we use positions instead of angles. In Fig. \ref{fig10} we observe that for $B_{att}=0$ and $B_{att}=4$ for $\omega=22.5^{\circ}$ at $\phi=0.02$ only two angles are most probable $20^{\circ}$ and $160^{\circ}$. We already know that $Z_{PI}<2$ which means that on average we form one bond per patch or in other words we have chain conformation. We also observe that around the angle of $40^{\circ}$, $g(\theta)$ converges to zero faster for $B_{att}=0$ and $B_{att}=4$ compared to $B_{att}=12$. A tale is devolped for the distribution of $B_{att}=12$ close to $40^{\circ}$ and $140^{\circ}$ which is the signature of bundling at very low concentration. Tavares et al. \cite{tavares2012quantitative} have shown that for the case of $B_{att}=0$ the most probable angle that they observe for the case of $22.5^{\circ}$ turns out to be $22^{\circ}$ which agrees with our results as well. 

In Fig. \ref{fig11} we have plotted the distribution of $g(\theta)$ for three different $\phi$ and $B_{att}$ values at $\omega=22.5^{\circ}$. Here we observe that even for $B_{att}=0$ all the angles between $40^{\circ}$ and $140^{\circ}$ are possible for moderate to high volume fractions. As the volume fraction increases the rearrangement of the patches as well as the diffusion of the cluster for aggregation becomes difficult, as it may lead to bond breakage or may lead to overlap with the neighboring spheres. Also the maximum for the most probable angle that we observed for low volume fraction comes down and shifts inwards for moderate and high $\phi$ values as the particles maximize the reversible bonds thus going to a lower energy state. In this case we also observe another peak which appears at $90^{\circ}$ which means the particles form bundles and these bundles branch out similar to the spaghetti like structure.  

In the present study even though multiple bonds are allowed for the patches, $\omega=22.5^{\circ}$ was able to form only one bond per patch.  For the case of $\omega=45^{\circ}$ as shown in Fig. \ref{fig12} we observe that peaks which existed for the case of $\omega=22.5^{\circ}$ are not very prominent and that the maximum has shifted to $60^{\circ}$ and $120^{\circ}$ for all the  $B_{att}$ and $\phi$ values. This can be understood from the fact that for $\omega=45^{\circ}$ we are allowing the patches to form multiple bonds. The patchy particles will try to maximize the irreversible bonds so the most probable angle change from the case of $\omega=22.5^{\circ}$. It has already been shown that the average number of bonded particles per patch increases from $2$ to $4$ see Fig. \ref{fig5}. As the attraction increases there is another probable angle coming up at $90^{\circ}$ for all the volume fractions we have studied. This is due to the the reversible part of the potential, as for $B_{att}=0$ (also see Fig. \ref{fig11}) for all the volume fraction we have considered in the present study we observe a minimum at $90^{\circ}$. Thereby we can conclude that reversible part of the potential favors branching of the clusters as observed for the case of $\omega=22.5^{\circ}$.  

In the present work we have shown that for the case of $\omega=22.5^{\circ}$, the patchy particles always form chains for $B_{att}=0$ and $B_{att}=4$. Whereas at $B_{att}=12$ we observe that chains aggregate together to form bundles. It was suggested by Huisman et al.\cite{huisman2008phase} that this transformation is similar to the sublimation transition of polymers. They have shown that the transition happens at a specific temperature using Monte-Carlo simulations. In the present work we also observe a very sharp transition from chains to bundles. The kinetics of aggregation in the case of $NPI$ cluster construction gives us a clear indication that the transition is similar to a nucleation and growth type mechanism. We also observe that the chain length at $B_{att}=4$ where only chains are formed, is higher than for the case of $B_{att}=12$ where bundling happens. This is contrary to the work of Huisman et al. \cite{huisman2008phase} , because in their case the individual particle can break and form bonds, while in the present study we have irreversible bonds for the patches. So once the bonds are formed among the patch they grow as chains for the case of $\omega=22.5^{\circ}$ and these chain aggregate together to form bundles. In order for the chain to grow they have to align themselves and the probability for $2$ chains to align is very small, due to the hindrance created by other cluster around. 

The chain to bundle transition is also believed to be the reason for many neurodegenerative diseases like alzheimer's. There the monomers aggregate to form oligomers which then transfer to bundle configuration commonly called amyloid fibers. These fibers are chemically stable quite similar to our present model where irreversible bonds give structures formed a stable conformation. These fibers then aggregate together to form a percolating cluster or gels depending on the PH of the solution, which is very similar to the present model where we started with monomer which aggregate together to form chains which transformed into a bundle depending on the interaction strength or the quality of the solvent. These amyloid fibers form helical bundles similar to the present study for individual arm of the gel formed for the case of $\phi=0.02$, $\omega=22.5^{\circ}$ at $B_{att}=12$ see figure \ref{fig2}(c). The present model gives a good qualitative description of the different process involved in the amyloid fiber formation more quantitative study will be reported in the future.

\section{Conclusions}\label{conclu}
We have used Brownian Cluster Dynamics to investigate the kinetics of aggregation and structure of the resulting aggregates for the $2$ patch system by varying patch size and solvent condition. In the model studied, the anisotropic potential is complemented by square well isotropic potential, where b we have also defined $3$ different definition of clusters. We have  observed that for small patches we form chain like configuration and also the average chain length increases on increasing the isotropic interaction from  $B_{att}$=0 to  $B_{att}$=4. For $B_{att}$=12 the average chain length is less than  $B_{att}$=4 as a result of the chain to bundle transition via nucleation and growth mechanism. When the size of the patch was increased to $\omega=45^{\circ}$ we observed globular structure instead of the chains like configuration. We have shown that in the case of small patch size ($\omega=22.5^{\circ}$) the structure and kinetics of aggregation is dominated by the isotropic reversible interaction while for larger patch size ($\omega=45^{\circ}$) it is dominated by irreversible patchy interaction and on further increasing the patch size the system will tend towards the isotropic irreversible DLCA aggregation. In the case of $\omega=22.5^{\circ}$ the bond distribution around a single particle showed that only $2$ bond angles were most probable indicating chain formation. On further increasing the attraction the distribution developed a tail indicating chain to bundle transition. While for the case of $\omega=45^{\circ}$ we observed that there is a contribution from other angles as well giving us a more globular conformation.  It will be interesting to study how the kinetics of aggregation and structure of aggregates changes on making the anisotropic interactions also reversible which will be addressed in future work.

\section{Acknowledge}
We would like to thank the HPC Padum and Badal of IIT Delhi for providing us the necessary computational resources.

\end{document}